\documentclass[aps, pra, twocolumn, superscriptaddress, amsmath, 
tightenlines, longbibliography]{revtex4-2}

\usepackage{amssymb}
\usepackage{amsmath}
\usepackage{dcolumn}
\usepackage{graphicx}
\usepackage{mathrsfs}
\usepackage{appendix}
\usepackage{graphicx}
\usepackage{booktabs}
\usepackage{units}
\usepackage{color}

\def \hide#1{}

\usepackage{url}
\usepackage[colorlinks]{hyperref}
\hypersetup{%
	plainpages=true,
	breaklinks=true,       
	hypertexnames=false,  
	pageanchor=true,
	colorlinks=true,
	linkcolor={blue},
	citecolor={red},
	urlcolor={blue},
	anchorcolor={black}
}

\begin{document}

\title{Long-Range Four-body Interactions in Structured  Nonlinear Photonic 
Waveguides}

\author{Xin Wang}
\affiliation{Institute of Theoretical Physics, School of Physics, Xi'an Jiaotong 
University, Xi'an 710049, People’s Republic of China}
\affiliation{Quantum Computing Center, RIKEN, Wakoshi, Saitama 351-0198, 
Japan}

\author{Jia-Qi Li}
\affiliation{Institute of Theoretical Physics, School of Physics, Xi'an Jiaotong 
University, 	Xi'an 710049, People’s Republic of China}

\author{Tao Liu}
\email{liutao0716@scut.edu.cn}
\affiliation{School of Physics and Optoelectronics, South China University of 
Technology, Guangzhou 510640, China}

\author{Adam Miranowicz}
\affiliation{Institute of Spintronics and Quantum Information,
Faculty of Physics, Adam Mickiewicz University, 61-614 Pozna\'{n}, 
Poland}
\affiliation{Quantum Computing Center, RIKEN, Wakoshi, Saitama 351-0198, 
	Japan}

\author{Franco Nori}
\affiliation{Quantum Computing Center, RIKEN, Wakoshi, Saitama 351-0198, 
	Japan}
\affiliation{Physics Department, The University of Michigan, Ann Arbor, Michigan 
48109-1040, USA}
\date{\today}

	\begin{abstract}		
		Multi-photon dynamics beyond linear optical materials are of significant fundamental and technological importance in quantum information processing. However, it remains largely unexplored in nonlinear waveguide QED.		
		In this work, we theoretically propose a structured nonlinear 
		waveguide in the presence of staggered photon-photon interactions, 
		which supports two branches of gaped bands for doublons (i.e., 
		spatially 
		bound-photon-pair states). In contrast to linear waveguide QED 
		systems, 
		we identify two important contributions to its dynamical evolution, 
		i.e.,  single-photon bound states (SPBSs) and doublon bound states 
		(DBSs). Most remarkably, the nonlinear waveguide can mediate the 
		long-range four-body interactions between two emitter pairs,  even 
		in 
		the presence of disturbance from SPBSs. By appropriately designing 
		system's 
		parameters, we can achieve high-fidelity four-body Rabi oscillations 
		mediated only by virtual 
		doublons in DBSs. Our findings pave the way for applying structured 
		nonlinear waveguide QED in multi-body quantum information processing 
		and quantum 
		  simulations 
		  among remote	sites.
	\end{abstract}
	
	\maketitle
	
\section{Introduction}
The past few years have witnessed a surge of 
interest in the field of  waveguide quantum electrodynamics (QED) in 
structured linear optical materials  without photon-photon interactions,  
leading to   intriguing phenomena  such as unconventional bound states, 
non-Markovian evolution and chiral 
emissions~\cite{Mitsch2014,Lodahl2015,Douglas2015,Bliokh2015a,Shi2016,Lodahl2017,
	GonzlezTudela2017,Liu2017,Trainiti2016,Chang2018,Elcan2020,Tang2022,Wang2022,Stewart2022,Bernardis2023}.
	 In these linear optical waveguides, dynamical properties are governed 
	and investigated at the single-photon level 
	\cite{Karplus1955,Iacopini1979,Scully1997,cohen1998atom,Walls07,Zhou2008,
		Liao2010b,Agarwal2012}. However, once quantum many-body interactions 
		between individual 
		photons are introduced, standard descriptions based on single-photon
	properties are 
	inadequate~\cite{Imamo1997,Liao2010,Peyronel2012,Dudin2012,Chang2014,Dorfman2016,
		Roy2017,Kruk2018,Sahand2018,Ke2019,Poshakinskiy2021,
		Marques2021,Sheremet2023}. These nonlinear quantum optics phenomena 
		can find important applications in fields of quantum information and 
	metrology~\cite{Napolitano2011,Reiserer2013,Shomroni2014,Mahmoodian2018,Bin2020,Cordier2023}.

In artificial platforms, such as nanophotonic structures and 
	circuit-QED, 
	strong nonlinear interactions can be experimentally 
	realized~\cite{Cramer2008,Ma2019,Eckardt2017,Rubio2020,Kuo2020,Zhu2022}, 
	providing ideal platforms  for exploring 
	quantum effects at the level of few 
	photons, and effects of many-body 
		statistics on waveguide 
		QED~\cite{Gemelke2009,Macha2014,Wei2015,Kaufman2016,Salath2015,
			Reithmaier2015,Roushan2017,Yan2019,ChangC2020,Carusotto2020,
			Kim2021,Scigliuzzo2022,Qiao2022,Zhang2023}.
			Recently, a remarkable supercorrelated 
		radiance phenomenon, beyond the conventional super- and subradiance, 
		was 
		reported in a nonlinear waveguide QED system with 
		non-structured bath~\cite{Wangz2020L}. Until now, in spite of 
		potentially	interesting physics hidden behind it, the field of 
		nonlinear 
		waveguide QED remains largely unexplored. Along previous works in 
		this field \cite{Wangz2020L, Talukdar2022,Talukdar2022a,Talukdar2023}, a natural question 
		arises: how the 
		structured nonlinear waveguide QED system influences the dynamics in 
		the 
		presence of a strong photon-photon interaction.
		
Here we consider emitter pairs interacting with a structured nonlinear 
waveguide by designing a 
staggered onsite photon-photon interaction. In contrast to previous work 
\cite{Wangz2020L, Talukdar2022,Talukdar2022a,Talukdar2023}, we find the 
formation of gaped 
bands for doublons, i.e., spatially bound-photon-pair 
states~\cite{Winkler2006,Piil2007,Wang2010,Liberto2016,
	Gorlach2017,Calaj2016,Tai2017,Lyubarov2019,ChenJD2020,
	Flannigan2020,Stepanenko2020,Xing2021,Berti2022,Stepanenko2022}.  
	We 
reveal the effects 
of both doublon 
bound states (DBSs)	and	single-photon bound states (SPBSs) on the 
evolution dynamics of the hybrid system. Most remarkably, we demonstrate
\emph{long-range four-body interactions}, mediated by DBSs, between 
distant emitter pairs. We find that a high-fidelity 
four-body interaction requires two emitter pairs for a  separation larger 
than 
the size 
of the SPBS, to prevent undesired single-photon-mediated
transitions. Our study shows a remarkable different physics 
in the nonlinear 
waveguide QED regime in contrast with conventional linear systems.
	
\section{Doublon energy bands}
\begin{figure*}[tbph]
	\centering \includegraphics[width=14cm]{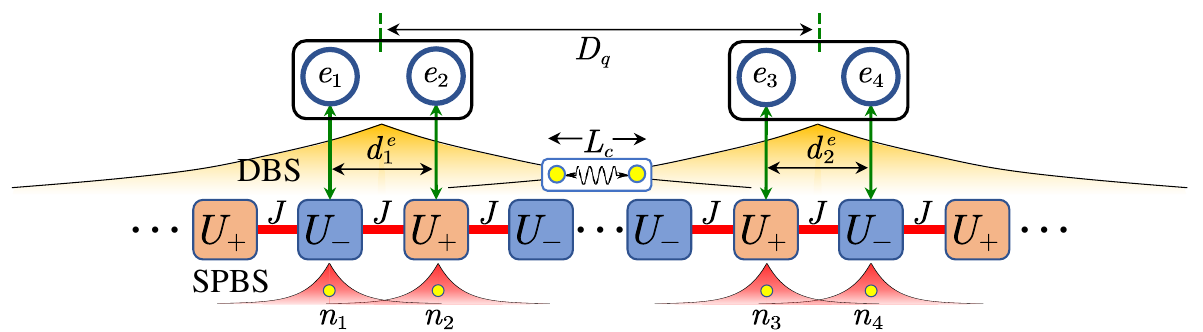}
	\caption{Schematic  of the nonlinear QED setup, with a waveguide 
		consisting of
		coupled cavity arrays in the presence of a Kerr nonlinearity. Here, 
		$J$ 
		is the photonic hopping strength, and $U_{\pm}=U_c\pm U_m$ denotes 
		the 
		staggered onsite photon-photon interaction. The emitter pair 
		$e_{1,2}$ is 
		separated from the pair $e_{3,4}$ by a distance $D_q$, while 
		$d_{1}^e$  
		and $d_{2}^e$ are the distance of each emitter pair. In the dynamical 
		evolution, a doublon bound state (DBS), corresponding to a virtual 
		exchange process of two photons with correlated length $L_c$, is 
		considered, accompanied by single-photon bound state (SPBSs) due to 
		the 
		virtual exchange of a single photon.}
	\label{fig1m}
\end{figure*}

\begin{figure}[tbph]
	\centering \includegraphics[width=8.2cm]{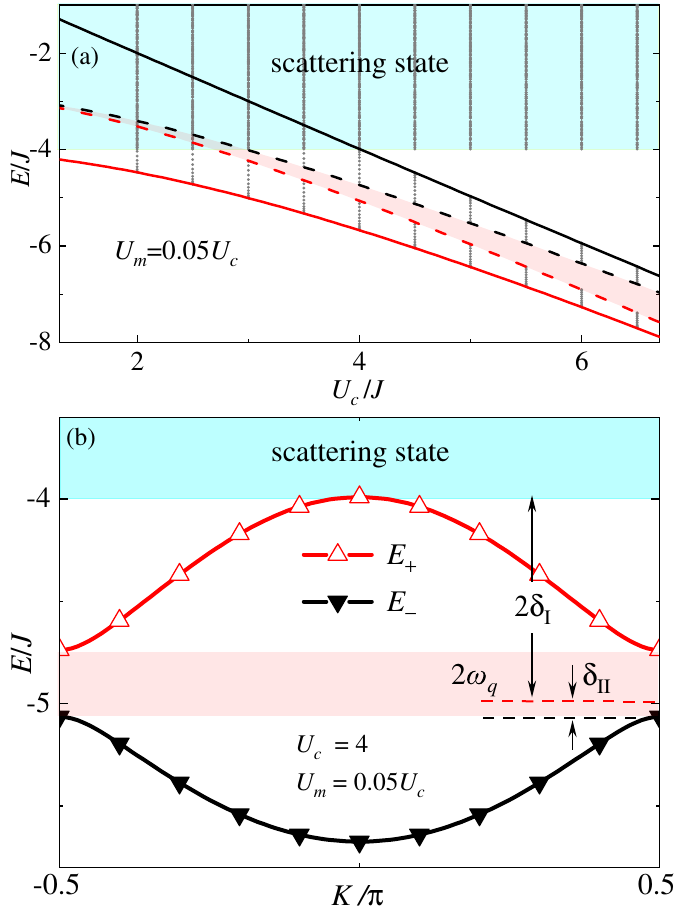}
	\caption{(a) Two-excitation spectrum of the Hamiltonian $H_w$
		versus the mean value $U_c$. The 
		curves represent the 
		upper and lower bounds of the doublon energy bands in 
		Eq.~(\ref{dis_K}). (b) 
		Doublon's spectrum for $U_c=4J$ and
		$U_m=0.05U_c$. The emitter-pair frequency $2\omega_q$ lies inside 
		the doublon band gap 
		with a detuning $\delta_{\mathrm{II}}$ 
		($2\delta_{\mathrm{I}}$) to the lower band (scattering 
		states). The cyan (pink) region corresponds to the scattering states (doublon band gap). }
	\label{fig2m}
\end{figure}

We consider a nonlinear waveguide 
constructed by an array of coupled cavities 
in the presence of strong 
photon-photon interactions (see Fig.~\ref{fig1m}).
In the rotating frame 
of the cavity frequency, the Hamiltonian of the waveguide is written as
\begin{equation}
H_w=-J\sum_n (a^\dagger_n 
	a_{n+1}+\text{H.c.})-\frac{1}{2}\sum_n U_n a^\dagger_n  
	a^\dagger_n 
	a_n a_n,
	\label{H_c}
\end{equation}
where $J$ is the bosonic
hopping strength, and $U_n$ denotes the Kerr 
nonlinearity at site $n$, and $a_{n}^{\dagger}$ is the 
annihilation operator of $n$th cavity. To realize a structured environment, 
we introduce 
the staggered photon-photon 
interaction with $U_n=U_c+(-1)^n U_m$ (where $U_c\pm U_m$ is the staggered 
interaction strength). We numerically calculate the 
two-photon spectrum as a 
function of $U_c$ for  $N=100$ with periodic boundary condition (by setting 
$J=1$), as shown in Fig.~\ref{fig2m}(a).  
Besides continuum scattering 
states, there exist discrete doublon energy bands  with  bound photon pairs. In addition, the doublon bands become  distinctly 
separated from the scattering states for  $U_c>4J$. 

The doublon spectrum 
can  be analytically solved by defining the center-of-mass and relative 
coordinates, i.e., $x_c=(m+n)/2$ and  $r=m-n$,  with $m$ and $n$ representing 
the positions of the two photons. 
Given that the 
Hilbert space is confined to a two-photon subspace, the nonlinear term 
can be expressed as
\begin{equation}
H_U=V(x_c,r)=[ U_c+U_m\cos \left(\pi x_c \right) ] \delta _{r0},
\end{equation}
where $\delta _{r0}$ is a delta function that is non-zero solely when 
$r=0$, and $x_c$ is restricted to be integers. It is important to mention 
that 
$H_U$ is modulated periodically in 
the 
$x_c$ direction. Applying the Bloch theorem, the eigenwavefunction of the 
doublon 
can be
written as
\begin{equation}
\Psi \left( m,n \right) =\frac{1}{\sqrt{N }}\exp(iKx_c)u_K\left( x_c,r  
\right),
\end{equation}
where $u_K\left( x_c,r  
\right)$ is a periodic function satisfying $u_K\left( x_c,r  \right) 
=u_K\left( 
x_c+2, r  
\right)$. In our study, the nonlinearity varies between nearest-neighbor 
sites, which results in the Fourier series having only two terms in the 
$x_c$ direction. We utilize a wavefunction ansatz in a separable variable 
form,
\begin{equation}
u_K\left( x_c,r  \right)=\psi _{K}^{(0)}\left( r \right) +e^{i\pi x_c} \psi 
_{K}^{(1)}\left( r \right). \label{u_K}
\end{equation}

The dispersion relation and wavefunctions are derived by solving the Schrödinger equation
\begin{gather}
\left[ H_0+\delta _{r0}H_U \right] \left[ \begin{array}{c}
\psi _{K}^{(0)}\\
\psi _{K}^{(1)}\\
\end{array} \right] =E\left[\begin{array}{c}
\psi _{K}^{(0)}\\
\psi _{K}^{(1)}\\
\end{array} \right].
\label{static_Seq}
\end{gather}
First, we start from the non-perturbation part, ignoring $H_U$. The first term in Eq.~(\ref{u_K}) is 
\begin{flalign}
&H_0e^{iK\frac{m+n}{2}}  \psi _{K}^{(0)}\left( m-n \right) 
\notag \\
&= - J\sum_{\pm} \left[ e^{iK\frac{m\pm 1+n}{2}}\psi _{K}^{(0)}\left( m\pm 1-n \right) \right.
\notag \\
&\qquad \quad \quad \left. +e^{iK\frac{m+n\pm 1}{2}}\psi _{K}^{(0)}\left( m-n\mp 1 \right) \right], \quad
\end{flalign}
which can be simplified as 
\begin{equation}
H_0\psi _{K}^{(0)}\left( r \right) =-2J\cos \left( \frac{K}{2} \right) \sum_{\pm}{\psi _{K}^{(0)}\left( r\pm 1 \right)}.
\label{H_0psi0}
\end{equation}
Similarly, the second term in Eq.~(\ref{u_K}) is 
\begin{equation}
H_0\psi _{K}^{(1)}\left( r \right) =-2J\cos \left( \frac{K+\pi}{2} \right) \sum_{\pm}{\psi _{K}^{(1)}\left( r\pm 1 \right)}.
\label{H_0psi1}
\end{equation}
Then $H_0$ is rewritten as 
\begin{align}
H_0&=\left| \begin{matrix}
H_{00}&		H_{10}\\
H_{01}&		H_{11}\\
\end{matrix} \right| \notag \\
&=\left| \begin{matrix}
-2J\cos \left( \frac{K}{2} \right) \Delta ^{\dagger}\left( r \right)&		0\\
0&		-2J\cos \left( \frac{K+\pi}{2} \right) \Delta ^{\dagger}\left( r \right)\\
\end{matrix} \right|, 
\end{align}
with  $$\Delta ^{\dagger}\left( r \right) \psi _{K}^{i}\left( r \right) 
=\left[ \psi _{K}^{(i)}\left( r+1 \right) +\psi _{K}^{(i)}\left( r-1 \right) 
\right].$$ 
The photon-photon interaction term $H_U$ is written as 
\begin{equation}
H_U=\left| \begin{matrix}
U_{00}&		U_{10}\\
U_{01}&		U_{01}\\
\end{matrix} \right|,
\end{equation}
where the matrix elements are respectively derived as:
\begin{align}
U_{00}&=\frac{1}{N}\sum_{x_c\in \mathbb{Z}}{e^{-iKx_c}H_Ue^{iKx_c}}=U_c,
\\
U_{01}&=U_{10}^{*}=\frac{1}{N}\sum_{x_c\in \mathbb{Z}}{e^{-iKx_c}H_Ue^{iKx_c}}e^{i\pi x_c}=U_m,
\\
U_{11}&=\frac{1}{N}\sum_{x_c\in \mathbb{Z}}{e^{-iKx_c}e^{-i\pi x_c}H_Ue^{iKx_c}}e^{i\pi x_c}=U_c,
\end{align}
where $H_U=U_c+U_m\cos \left( \pi x_c \right) $. By defining the Green function as 
\begin{equation}
\left( E-H_0 \right) G_K(E,r)=\delta _{r0}, \label{EH0}
\end{equation}
we obtain the Lippmann-Schwinger equation for the doublon states
\begin{gather}
\Psi_K (r) =\Psi _0(r)+\int G_K(E,r-r')\;\delta _{r0}\;H_U\;\Psi (r')\;dr', 
\notag \\
\Psi_{K} (r)  =\left[ \begin{array}{c}
\psi _{K}^{(0)}(r)\\
\psi _{K}^{(1)}(r)\\
\end{array} \right],
\end{gather}
where $\Psi _0(r)$ is the solution satisfying the noninteracting Hamiltonian $H_0$.
Employing the properties of the $\delta$-function, we obtain 
\begin{equation}
\Psi_{K} (r)=\Psi _0(r)+G_{K}(E,r)H_U\Psi_{K} (r=0).
\label{PSIK}
\end{equation}
Finally, the wavefunction at $r=0$ is derived as 
\begin{equation}
\left[1-G_{K}(E,r=0)H_U\right]\Psi_{K} (r=0) =\Psi _0(r=0).
\label{psi00}
\end{equation}

The probability of the scattering state, where the two photons
are not bound together, is zero; which corresponds to
$\Psi_0(r)\equiv 0$. Consequently, the doublon state corresponds to the 
nontrivial solution 
of the linear homogeneous equations in Eq.~(\ref{psi00}), 
\begin{equation}
\det \left[ 1-G_K(E,r=0)H_U \right] =0,
\end{equation}
from which the dispersive relation of 
the doublon spectrum is obtained.
\begin{figure*}[tbph]
	\centering \includegraphics[width=17.7cm]{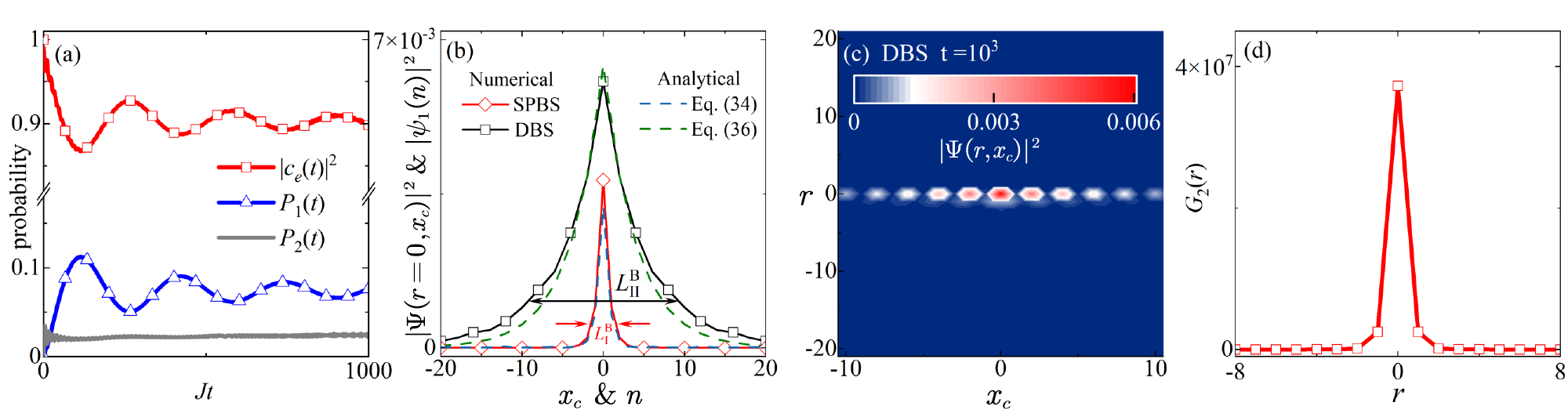}
	\caption{(a) Time-dependent probability  of the two-emitter excitations  
		$|c_e(t)|^2$, and single- and two-photon excitations $P_{1,2}(t)$  
		for 
		initially two-emitter excitation.  The 
		coupling 
		points are set at $n_{1,2}=0$. 
		(b) The squared modulus of wavefunction profile of the 
		SPBS and DBS (by restricting $r=0$) 
		obtained 
		via numerical simulations up to $t=10^3$. (c) 2D field 
		distribution for the DBS. 
		(d) Two-point 
		correlation function $G_2(r)$ for the field in (c). Here we 
		set 
		$J=1$, $g=0.1$, $U_c=4$, $U_m=0.05U_c$, $\alpha\simeq 2$, and 
		$\delta_{\mathrm{II}}=0.03$.}
	\label{fig3m}
\end{figure*}
The Green function in momentum space reads
\begin{equation}
G_K(E,r)=\frac{1}{2\pi}\int_{-\pi}^{\pi}{dq}\;G_K(E,q)\;e^{iqr}. \label{GKr0}
\end{equation}
By substituting Eq.~(\ref{GKr0}) into Eq.~(\ref{EH0}) and employing the 
properties of the 
$\delta$-function, we obtain 
\begin{align}
G_K(E,q)&=\frac{1}{\left( E-H_0 \right)}
\notag \\
&=\left| \begin{matrix}
\frac{1}{E+4J\cos \left( \frac{K}{2} \right) \cos q}&		0\\
0&		\frac{1}{E-4J\sin \left( \frac{K}{2} \right) \cos q}\\
\end{matrix} \right|.
\end{align}
Consequently, the Green function in real space is derived as
\begin{align}
G_K(E,r)&=\left| \begin{matrix}
f_{K}^{(0)}\left( E,r \right)&		0\\
0&		f_{K}^{(1)}\left( E,r \right)\\
\end{matrix} \right|,   \notag\\
f_{K}^{(0)}\left( E,r \right) &=\int_{-\pi}^{\pi}{\frac{e^{iqr}}{E+4J\cos 
		\left( \frac{K}{2} \right) \cos q}dq}, \notag \\
f_{K}^{(1)}\left( E,r \right) &=\int_{-\pi}^{\pi}{\frac{e^{iqr}}{E-4J\sin \left( \frac{K}{2} \right) \cos 
		q}dq}.
\label{FKpi}
\end{align}
By setting $r=0$, we write Eq.~(\ref{psi00}) as
\begin{equation}
1-G_K(E,r=0)H_U= \left| \begin{matrix}
\frac{U_c}{U_{\cos}}-1&		\frac{U_m}{U_{\cos}} \vspace{1ex}\\
\frac{U_m}{U_{\sin}}&		\frac{U_c}{U_{\sin}}-1 \\
\end{matrix} \right|,
\end{equation}
where 
\begin{gather*}
	U_{\cos}=\sqrt{E^2-16\left[ J\cos \left( K/2 \right) \right] ^2}, \\
	U_{\sin}=\sqrt{E^2-16\left[ J\sin \left( K/2 \right) \right] ^2}.
\end{gather*}

Altogether, the eigenenergy $E$ and the momentum of center-of-mass $K$ 
satisfy the following relationship
\begin{equation}
\det \left| \begin{matrix}
\frac{U_c}{U_{\cos}}-1&		\frac{U_m}{U_{\cos}} \vspace{1ex}\\
\frac{U_m}{U_{\sin}}&		\frac{U_c}{U_{\sin}}-1 \\
\end{matrix} \right|=0. 
\label{dis_K}
\end{equation}
We plot two branches of the doublon bands in Fig.~\ref{fig2m}(b) according to Eq.~(\ref{dis_K}), where a band gap exists.

Moreover, the amplitudes $\psi _{K}^{(1,2)}(r=0)$ correspond to the nontrivial solution 
of the linear homogeneous equations in Eq.~(\ref{psi00}). Therefore
\begin{gather}
\left[1-G_K(E,r=0)H_U \right]
\left[ \begin{array}{c}
\psi _{K}^{(0)}(r=0)\\
\psi _{K}^{(1)}(r=0)\\
\end{array} \right]
=\left[ \begin{array}{c}
0\\
0\\
\end{array} \right] \notag \\
\longrightarrow
\frac{\psi _{K}^{(0)}\left( r=0 \right)}{\psi _{K}^{(1)}\left( r=0 
\right)}=\frac{U_mU_{\sin}}{U_{\sin}U_{\cos}-U_cU_{\cos}}.
\label{fk01}
\end{gather}
By substituting Eq.~(\ref{fk01}) into Eq.~(\ref{PSIK}), we derive the 
wavefunctions for the doublon eigenstates as
\begin{align}
&\Psi (r)=\left[ \begin{array}{c}
\psi _{K}^{(0)}(r)\\
\psi _{K}^{(1)}(r)\\
\end{array} \right]=G(E,r)H_U\left[ \begin{array}{c}
\psi _{K}^{(0)}(r=0)\\
\psi _{K}^{(1)}(r=0)\\
\end{array} \right] \notag \\
&=
\left| \begin{matrix}
f_{K}^{(0)}\left( E,r \right)&		0\\
0&		f_{K}^{(1)}\left( E,r \right)\\
\end{matrix} \right|
\left| \begin{matrix}
U_c&		U_m\\
U_m&		U_c\\
\end{matrix} \right|\left[ \begin{array}{c}
\psi _{K}^{(0)}(r=0)\\
\psi _{K}^{(1)}(r=0)\\
\end{array} \right], \label{phiKr}
\end{align}
where we have employed the condition $\Psi _0(r)\equiv 0$. Via analysis the integral Eq.~(\ref{FKpi}), we obtain the formal solution of the wavefucntion
\begin{gather}
\Psi _K\left( x_c,r \right) \propto \exp \left[ -\frac{|r|}{L_c(K)} \right] ,
\end{gather}
which $L_c(K)$ is the correlated length of the photon pair. In particular, at the band edge with zero
group velocity, the wavefuntions $\Psi^\pm_{K}$ at $K_0=\pi/2$  of the upper and lower bands can be analytically obtained as (see Appendix \ref{A}) 
\begin{equation}
\Psi _{K_0}^{\mp}\left( x_c,r \right) =\frac{e^{iK_0x_c}\left( 1\pm e^{i\pi x_c} \right)}{\sqrt{N}\psi _0}\exp \left[ -\frac{|r|}{L_c^{\mp}(K_0)} \right], 
	\label{K_0_psi}
\end{equation}
where $\psi_0$ is the normalization factor, and the correlated length is  
$$L_c^{\mp}(K_0) = \left( \ln \frac{2\sqrt{2}J}{-E_{\mp}-\sqrt{E_{\mp}^{2}-8J^2}} \right)^{-1}.$$
A strong on-site nonlinear interaction  results in 
supercorrelated doublon modes. 
According to Eq.~(\ref{K_0_psi}),  $L_c^{\mp}(K_0)\sim 1$ for 
$U_c\simeq 4J$, indicating that the two photons are strongly bunched in 
space.

\section{Dynamics and bound states}
\label{III}
\subsection{Dynamics evolution}

We consider two emitters $e_1$ 
and $e_2$, which are separated by a distance $d^e_{1}$,  interacting with the 
nonlinear 
waveguide at points $n_{1 }$ and $n_{ 2}$ (see Fig.~\ref{fig1m}). The hybrid system's 
Hamiltonian is 
\begin{equation}
H=H_w+\frac{\omega _q}{2}\sum_{i=1}^2{\sigma _{i}^{z}}+g\sum_{i=1}^2{\left( \sigma 
_{i}^{+}a_{n_i}+\mathrm{H}.\mathrm{c}. \right)},
\label{Hint_m}
\end{equation}
where $a_{n_i}^{\dagger}$ is the annihilation operator of the
bosonic mode interacting with the $i$th emitter, $\omega _q$ is the emitter 
frequency, and $g$ is the coupling strength 
between each emitter and the waveguide. When the two-emitter excitation 
frequency $2\omega_q$ is set in the band gap between the
doublon bands [see Fig.~\ref{fig2m}(b)], the evolution becomes
highly 
non-Markovian owing to the van Hove singularity in the density 
of states~\cite{Douglas2015}, which is 
different from the case of unstructured nonlinear waveguides 
\cite{Wangz2020L}. 

We now proceed to calculate the dynamics in the double-excitation subspace. We assume two emitters  initially 
in their excited  states, and their frequencies are chosen to be close to   
the lower band edge, with frequency detuning 
$\delta_{\mathrm{II}}=2\omega_q-E_-({K_0})\simeq 0$ [see 
Fig.~\ref{fig2m}(b)]. Since $H$ in Eq.~(\ref{Hint_m}) conserves the  
excitation number, in the double-excitation subspace, the   state $ |\psi 
_2\left( t \right) \rangle$  is
\begin{align}\label{Tstates_m}
|\psi _2\left( t \right) \rangle &=c_e\left( t \right) |ee,\mathrm{vac}\rangle 
+\sum_K{c_K\left( t \right)}|gg,\Psi^-_{K}\rangle \notag \\
&+\sum_k{\left[ c_{1k}\left( t \right) |eg\rangle +c_{2k}\left( t \right) 
|ge\rangle \right]}a_{k}^{\dagger}|\mathrm{vac}\rangle,
\end{align}
where $c_e\left( t \right)$ [$c_K\left( t \right)$] is the probability
amplitude for two emitters (doublon states $|\Psi^-_K\rangle$) being excited,
$a_{k}^{\dagger}=\sum_n{e^{ikn}a_{n}^{\dagger}}/\sqrt{N}$ is the single-photon 
creation operator in momentum space, and $c_{ik}$ is the 
probability 
amplitude for the emitter $i$ and the mode $k$ being simultaneously excited. 
Because $\omega_q$ is significantly detuned from the scattering states (i.e., 
$\delta_{\mathrm{I}}=|\omega_q-2J|\gg g$), we have neglected their 
contributions in Eq.~(\ref{Tstates_m}), and the one-photon state can also be adiabatically eliminated. Hence, we obtain the evolution equation (more detain in Appendix \ref{B1})
\begin{align} 
i\dot{c}_e\left( t \right) &=-\frac{1}{\sqrt{N}}\sum_K{\mathcal{G} _K\left( n_1,n_2 \right)}c_K\left( t \right) , \label{coupled_eq1}
\\
i\dot{c}_K\left( t \right) &=\Delta _Kc_K-\frac{1}{\sqrt{N}}\mathcal{G} _{K}^{*}\left( n_1,n_2 \right) c_e\left( t \right),\label{coupled_eq2}
\end{align}
where the emitter pair couples to the doublon mode $K$ with an effective transition rate 
\begin{gather}
\mathcal{G} _K\left( n_1,n_2 \right) =g^2\sum_{i,j=1,2}^{i\ne j}{\sum_k{\frac{e^{ikn_j}}{\delta _k}}}M(K,k,n_i). \label{G_K_1}
\end{gather}
$M(K,k,n_i)=\langle k,n_i|\Psi _{K}^{-}\rangle$ denotes the process which annihilating one photon at position $n_i$ and another photon with mode $k$ to create a doublon mode $K$. Hence, we can simplify Eq.~(\ref{G_K_1}) as
\begin{gather}
\mathcal{G} _K\left( n_1,n_2 \right) \! \propto \! \sum_{i,j=1,2}^{i\ne j}{\sum_n{\!\Psi _{K}^{-}( \frac{n+n_j}{2},n_j-n ) \psi _i( n )}},
\end{gather}
which is proportional to the overlap between the doublon mode $\Psi^-_K$ and a SPBS $\psi _i\left( n \right)$. This quantity implies the following dual processes: the emitter $i(j)$ excites a 
SPBS, distributing in the waveguide around the site $n_i(n_j)$; meanwhile, 
the emitter $j(i)$ excites another 
photon at $n_j(n_i)$. The overlap between this two photon pairs and the doublon 
state $\Psi^-_K$ will excite the doublon mode $K$. As discussed in Appendix \ref{B2}, the formula of SPBS is 
\begin{gather}
\psi_{i}(n)\simeq A_s 
\exp\left(-\frac{|n-n_i|}{L_{\mathrm{I}}^{\text{B}}}\right), 
\notag \\
\frac{1}{L_{\mathrm{I}}^{\text{B}}}=\ln \frac{2J}{|\omega_q|-\sqrt{
		\omega_q^2-4J^2}}, \quad A_s=\frac{gJ}{\sqrt{ \omega_q^2-4J^2}},
\label{psi1_m}
\end{gather}
where $L_{\mathrm{I}}^{\text{B}}$ 
($A_s$) is the decay length (amplitude) of the SPBS. Note that, $\mathcal{G}_K$ can be simplified further as
\begin{gather}
\mathcal{G} _K\left( n_i,n_j \right) \! \propto \! \sum_n\exp \! \left[ -\frac{|n-n_j|}{L_c^-\left( K \right)} \right] \! \exp \! \left(\! -\frac{|n-n_i|}{L_{\mathrm{I}}^{\mathrm{B}}} \!\right),
\end{gather}
which is decided by the decay length of double mode and SPBS, i.e., $\{L_{c}^{-}(K), 
L_{\mathrm{I}}^{\text{B}}\}$~\cite{Wangz2020L}. If $|n_i-n_j|\gg 
\max \{L_{c}^{-}(K), L_{\mathrm{I}}^{\text{B}}\}$, $\mathcal{G}_K\left( n_1,n_2 \right)$  decrease to zero, which implied the emitter pair decoupling to the doublon mode. 

\subsection{Doublon bound state}

\begin{figure}[tbph]
	\centering \includegraphics[width=8cm]{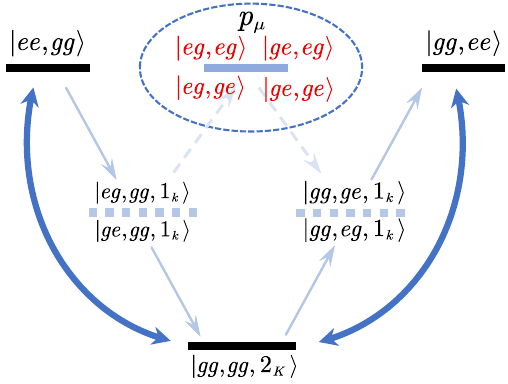}
	\caption{Illustration of the transitions mediated by the SPBSs and DBSs
		respectively. To enhance the fidelity of four-body Rabi 
		oscillations, the unwanted single-photon transition processes (dash lines) should be suppressed.}
	\label{fig4m}
\end{figure}

\begin{figure*}[tbph]
	\centering \includegraphics[width=17.8cm]{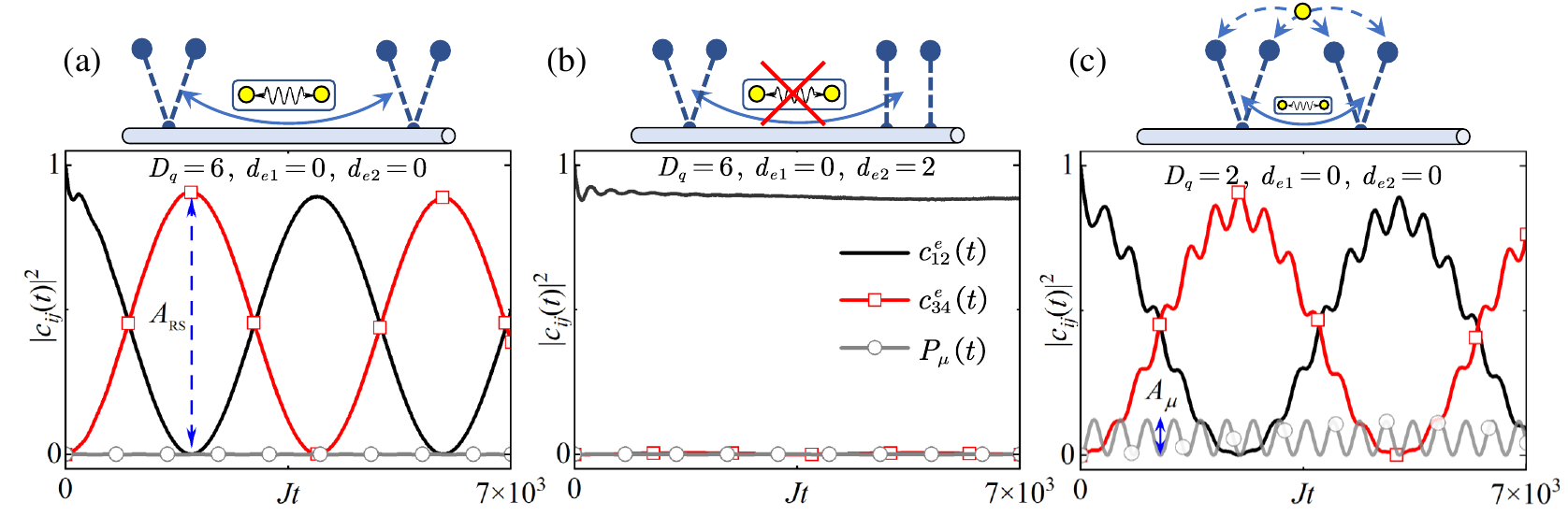}
	\caption{(a-c) Four-body Rabi oscillations 
		between two emitter pairs under different distance conditions. The 
		amplitude of the four-body Rabi oscillation 
		(single-photon 
		transitions) is denoted as $A_{\text{RS}}$ ($A_{\mu}$). Same 
		parameters as in 
		Fig.~\ref{fig3m}.
	}
	\label{fig5m}
\end{figure*}

As shown in Fig.~\ref{fig2m} the pink region, due to the nonlinear periodic 
structure $U_n=U_c+(-1)^nU_m$, energy gap emerges in the band. Basis on this 
band gap, we realize 
DBSs, a new bound states which contains two super-correlated photons. By 
setting the two-emitter excitation frequency $2\omega_{q}$ lies inside the 
doublon, the emitter pair is prevented from radiating doublons. Part of their 
excitation is trapped in the form of DBS.In 
Fig.~\ref{fig3m}(a), we plot
the dynamic evolution in a finite-size waveguide with $N=1000$, which is long 
enough to avoid the field being reflected by the open boundaries 
of the waveguide.
By approximating the dispersion relation around $E_-(K_0)$ as a quadratic form with curvature $\alpha$, i.e., $\Delta _K\simeq  \delta_{\mathrm{II}} 
+\alpha \left( K-K_0 \right) ^2$ , 
the analytical 
wavefunction of the DBS is derived as (more detail in Appendix \ref{B3})
\begin{gather}
\Psi _d(x_c,r)\! \simeq \! A_d\left(\! 1+e^{i\pi x_c} \! \right) \exp \left( \! -\frac{r}{L_c} \! \right) \exp \left( \! -\frac{|x_c-x_m|}{L_{\mathrm{II}}^{\mathrm{B}}} \! \right) ,
\notag \\
A_d=\frac{\mathcal{G}_{K_0}^{*}\left( n_1,n_2 \right)}{\psi _0\sqrt{\delta 
_{\mathrm{II}}\alpha}},\quad 
L_{\mathrm{II}}^{\mathrm{B}}=\sqrt{\frac{\alpha}{\delta _{\mathrm{II}}}},
	\label{DBS_state_m}
\end{gather}
where $x_m=n_1+n_2$. It shows that the decay length $L_{\mathrm{II}}^{\text{B}}$ of the DBS is 
determined by the detuning $\delta_\mathrm{II}$ and the band curvature. Given the proximity of $2\omega_q$ to the band edge, the stationary DBS 
can extend a considerable distance from the coupling points. Because only the modes around $K_0$ are 
excited 
with high probabilities, the correlation length is approximated as
$L_c^-(K_0)$. Note that, due to the DBSs, a new decay length $L_{\mathrm{II}}^{\mathrm{B}}$ emerges in the $x_c$ direction.


In Fig.~\ref{fig3m}(b, c), we plot the squared 
modulus of long-time 
field distribution
for single- and two-photon states, corresponding 
to SPBS and DBS described in Eq.~(\ref{psi1_m},\ref{DBS_state_m}). Owing to $L_{\mathrm{I}}^{\mathrm{B}}$ and $L_{\mathrm{II}}^{\mathrm{B}}$ is inversely proportional to $\delta_{\mathrm{I}}$ and $\delta_{\mathrm{II}}$, respectively. Hence, under the setup of Fig.~(\ref{fig2m}), both the amplitude and decay 
length of the DBS can be considerably larger than 
for the SPBS, i.e., $A_d > A_s$ and $L_{\mathrm{II}}^{\mathrm{B}} \gg L_{\mathrm{I}}^{\mathrm{B}}$. This is also manifested in Fig.~\ref{fig3m}(a), 
where the single-photon probability $P_1=\sum_{i,k} |c_{ik}(t)|^2$,   which is obtained numerically, is much lower than the two-photon probability $P_2= \sum_K |c_K(t)|^2$ after a long-time evolution. In the DBS, the two virtual photon are strongly bound with a correlation length $L_c^-(K_0)\simeq 1$, as shown in the $r$ direction of Fig.~\ref{fig3m}(c). To 
quantify this scale, we introduce the two-point correlation function for the 
DBS
\begin{equation}
G_2\left( r \right) =\sum_n{\frac{\langle 
		a_{n}^{\dagger}a_{n+r}^{\dagger}a_{n+r}a_n \rangle 
	}{\langle 
		a_{n+r}^{\dagger}a_{n+r}\rangle \langle a_{n}^{\dagger}a_n\rangle}}.
\end{equation}
As shown in Fig.~\ref{fig3m}(d), the decay length of $G_2\left( r \right)$ is 
of the same order as $L_c^-(K_0)\simeq 1$, and the two photons are strongly 
bunched in space.
Therefore, we realize two photons bound state with super-correlated photon 
pair, which two photons can't be separated and are jointly located around the 
coupling points. 

\section{Four-body interactions}
Via the nonlinear potential, the waveguide can mediate four-body interactions. To demonstrate this, we consider two emitter pairs coupled to the common nonlinear waveguide, as depicted in Fig.~\ref{fig1m}. 
The separation distance between the center of each pair is  set as $D_q$. In 
the nonlinear waveguide, there are two kinds of virtual processes due to the 
wavefunction 
overlaps of DBSs for two emitter pairs, and SPBSs 
for emitters in different pairs. 
For the former case, the virtual exchange between two-doublon states leads to 
a \emph{ four-body interaction } associated with the transition $|ee,gg\rangle 
\leftrightarrow |gg,ee\rangle$, as shown in Fig.~\ref{fig4m}. The four-body 
transition is a band-gap interaction which can 
	avoid dissipations led by the doublon continuous modes.

For the latter case, the virtual exchange of a single photon induced a 
conventional two-body interaction. For instance, as shown in 
Fig.~\ref{fig4m}, the overlap between two SPBSs of emitters 1 and 3 cause a 
two-body interaction associated with the transition $|ee,gg\rangle$ and 
$|ge,eg\rangle$.  There exist four distinct 
single-photon transition paths (see Fig.~\ref{fig4m}), and the total oscillating amplitude is
$P_\mu=\sum_{ij\neq 12,34}|c_{ij}^e(t)|^2$, where $c_{ij}^e(t)$ denotes the 
probability for emitters, labelled by  $i,j $, in the excited 
states.

For realizing high-fidelity four-body Rabi oscillations, the distance $D_q$ 
is set as $\gg\!\!L_{\mathrm{I}}^{\text{B}}$, which prevents any overlap 
between the SPBSs of emitters in different pairs and suppress the 
single-photon transition. Under the appropriate parameters, $P_\mu \simeq 0$, 
as shown in Fig.~\ref{fig5m}(a).
\begin{figure*}[tbph]
	\centering \includegraphics[width=17.5cm]{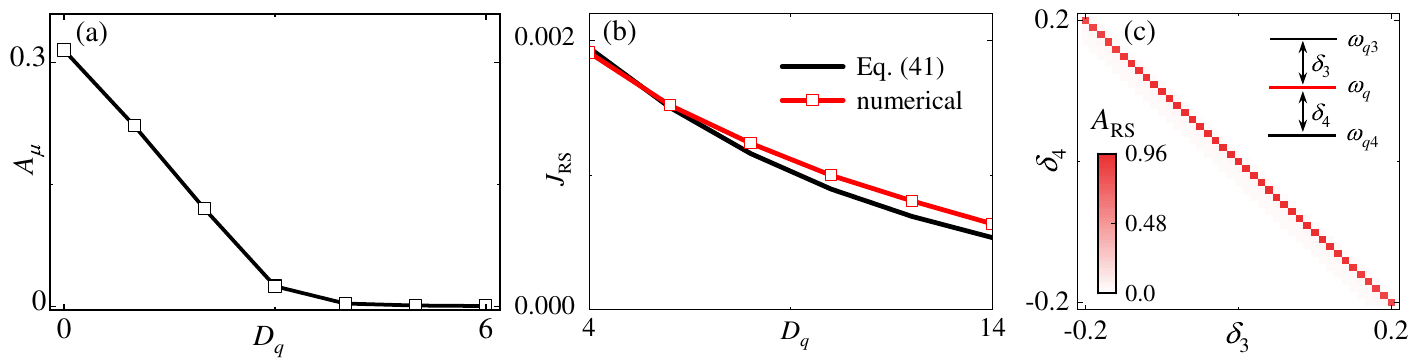}
	\caption{(a)The 
		single-photon transition amplitude $A_\mu$ versus $D_q$. (b)The four-body interaction strength $J_{\text{RS}}$ 
		versus
		$D_q$. The curve marked with symbols is obtained through 
		numerical simulation, while the solid curve is 
		plotted according to Eq.~(\ref{JS}). (c)Amplitude of the four-body Rabi oscillation 
		$A_{\text{RS}}$ versus 
		$\delta_{3,4}$. Same parameters	as in 
		Fig.~\ref{fig4m}(a).}
	\label{fig6m}
\end{figure*}
With $P_\mu \simeq 0$, the system can be described by an effective four-body interaction Hamiltonian 
\begin{align}
H_{\text{RS}}=&\Delta _{S1}\sigma _{1}^{+}\sigma _{2}^{+}\sigma _{1}^{-}\sigma 
_{2}^{-}+\Delta _{S2}\sigma 
_{3}^{+}\sigma _{4}^{+}\sigma _{3}^{-}\sigma 
_{4}^{-} \notag \\
&+\left(J_{\text{RS}}\sigma _{1}^{-}\sigma 
_{2}^{-}\sigma _{3}^{+}\sigma _{4}^{+}+\text{H.c.}\right), \label{Heff}
\end{align}
where
\begin{gather}
\Delta _{Si}\!=\!\frac{1}{N}\sum_K{\frac{|\mathcal{G}_{iK}|^2}{\Delta 
_K}},\quad  J_{\mathrm{RS}}\!=\!-\frac{1}{\pi}\! \int_0^{\pi}\! 
{\frac{\mathcal{G}_{1K}\mathcal{G}_{2K}^{*}}{\Delta _K}dK},
\\
\mathcal{G}_{1K}=\mathcal{G}_K\left( n_1,n_2 \right),\quad 
\mathcal{G}_{2K}=\mathcal{G}_K\left( n_3,n_4 \right) .
\end{gather}
The first two term in Eqs.~(\ref{Heff}) correspond to 
dynamical Stark shifts, and $\Delta _{S1}=\Delta _{S2}$ given that the frequencies of 
two pairs are identical. The latter terms denote the interaction between emitter pairs, and $J_{\text{RS}}$ is the effective four-body Rabi oscillation rate. Similar to the discussion for the DBS, only the modes around $K_0=\pm \pi/2$ are excited with high probabilities. Therefore, we can simplify $J_{\text{SR}}$ as
\begin{equation}
J_{RS}\simeq \frac{\mathcal{G} _{K_0}(n_1,n_2)\mathcal{G} _{K_0}^{*}(n_3,n_4)}{\sqrt{\delta _{\mathrm{II}}\alpha}}\exp \left( -\frac{D_q}{L_{\mathrm{II}}^{\mathrm{B}}} \right) , \label{JS} 
\end{equation}
which exponentially decreases as  
$D_q$, the same decay length as the DBS. In addition, emitters within the same pair are required to emit or absorb virtual photons concurrently, i.e., $\mathcal{G}_{iK}\ne 0$. Therefore, $d^e_{1,2}$ should be 
considerably smaller than the correlation length $\max \{L_c^{-}(K_0), 
L_{\mathrm{I}}^{\text{B}}\}$. Eventually, we summarize 
the parameter regimes, where the four-body Rabi oscillations occur with a high fidelity:
\begin{eqnarray}
\begin{cases}
1. \qquad  d^e_{1,2} \ \leq\  \max \{L_c^{-}(K_0), 
L_{\mathrm{I}}^{\text{B}}\},\\
2. \qquad  L_{\mathrm{I}}^{\text{B}}\ <\ D_q\ \sim\  
L_{\mathrm{II}}^{\text{B}}.
\end{cases}\label{consc}
\end{eqnarray}

By considering the parameters in Fig.~\ref{fig3m}, the length scales can be 
computed as follows: $L_{\mathrm{II}}^{\text{B}}\simeq 9$, 
$L_{\mathrm{I}}^{\text{B}}\simeq 1.6$, and 
$L_c^-(K_0)\simeq 1.4$. When $D_q$ and $d_{1,2}^e$ 
satisfy the conditions in 
Eq.~(\ref{consc}), the four-body Rabi oscillation happens with a high 
fidelity, which amplitude is $A_{\text{RS}}\simeq 1$,
as shown in Fig.~\ref{fig5m}(a). 
Once $d_{2}^e>2$, i.e., $\mathcal{G}_{2K}=0$, the second emitter pair 
decouple to the doublon mode, which cannot simultaneously absorb two virtual 
photons in the DBS. Consequently, the exchange process vanishes [see 
Fig.~\ref{fig5m}(b)]. Specifically, it is crucial to maintain a 
certain separation between the two pairs of emitters. When $D_q \approx 
L_{\mathrm{I}}^{\text{B}}$, the undesired transitions mediated by the SPBSs 
will disrupt the four-body Rabi oscillations, and the single-photon 
transition amplitude $A_\mu$ cannot be neglected, which can be confirmed by the 
evolution in Fig.~\ref{fig5m}(c). In Fig.~\ref{fig6m}(a, b), we plot $A_\mu$ 
and $J_{\text{RS}}$ versus $D_q$. When two emitter pairs are separated 
$D_q>4$, the single-photon transitions are super-suppress, and the SPBSs are 
negligible $A_\mu\simeq 0$. Moreover, owing to the van Hove singularity at 
doublon band edges, the DBSs can distribute over a distance of tens of unit 
cells, and the four-body Rabi oscillation occurs when two emitter pairs are 
separated a long distance. Even when $D_q>10$, $J_{\text{RS}}$ is nonzero 
[see Fig.~\ref{fig6m}(b)]. 

When the coupling layout of second pair is changed, the four-body interaction vanishes. However, this four-body interaction has a special robustness against the frequency shift. In Fig.~\ref{fig6m}(c), we plot $A_{\text{RS}}$ versus the frequency detuning 
of the second pair, i.e.,  $\delta_{3,4}=\omega_{3,4}-\omega_q$ by fixing the 
frequency of the first pair as $\omega_{q1,q2}=\omega_q$. 
Notably, when the summation frequency of the pair is fixed (i.e., 
$\delta_{3}+\delta_{4}=0$), the detuning of each emitters hardly affects the 
four-body Rabi oscillation. For detunings 
	$\{\delta_{3},\delta_{4}\}$ far away from the anti-correaltion line 
	$\delta_{3}+\delta_{4}=0$, the oscillation amplitude  $A_{\text{RS}}$ 
	quickly 
	vanishes to zero.
	Those phenomena indicate that, two photons are jointly emitted or 
	absorbed, and 
	behave as a single quasiparticle. Therefore, we realize the four-body 
	Rabi oscillations between two emitter pairs via the doublon bound state. 
	With appropriate parameters, the fidelity of the oscillations can reach 
	$100\%$ and the distance between two pairs can reach tens of unit cells 
	with still maintaining a high-fidelity. Therefore, our proposal exhibits 
	the potential to realize multi-body interactions between distant sites.

\section{Conclusion}

Here we have shown that both doublon 
bound states (DBSs)  and single-photon bound states (SPBSs) can be observed in a hybrid system of nonlinear 
waveguide and emitter pairs.  By appropriately tuning the system parameters, 
we can control the relative amplitude, for contributing to the system 
dynamics, of DBS and SPBS. Moreover,  a simplified form of the 
four-body 
interaction can only  be realized via  mediation of DBSs. We  analyze the 
conditions for realizing  high-fidelity four-body Rabi oscillations between 
two 
remote emitter-pairs.
Our proposal can be extended to a 
lattice chain with emitter pairs hopping together via the four-body 
interactions, which have potential applications in, for example,
lattice-gauge theory
simulations~\cite{Marcos2013,Brennen2016}.

The coupled transmon array, which
	has been experimentally studied in quantum 
	simulations~\cite{Ye2019,Olli2022,Xiang2023}, can be 
	configured
	as a nonlinear waveguide in this study. The hopping strength $J$ 
	in circuit-QED can be engineering into strong coupling regime with $J 
	\simeq 
	500~\text{MHz}$~\cite{Bosman2017,Kockum2019b_r,Ding2023}. As indicated in 
	Fig.4, the four-body 
	interaction 
	strength 
	can be around $J_{\text{SR}}\simeq 0.5\sim1\!~\:\text{MHz}$. In 
		current circuit-QED experimental setups, the intrinsic dissipation 
		rate 
		of an 
		individual transmon is  around $\gamma/(2\pi)\simeq 
		5\text{KHz}$~\cite{Krinner2022,Place2021}, which 
		is much weaker than $J_{RS}$. 
		Additionally, theoretical study in Ref.~\cite{Wangz2020L} also 
		showed the slow 
		dissipation 
		of the waveguide has little effect on the quantum phenomena with 
		doublons. Therefore, we believe that the predicted four-body 
		interaction is possible to observe in circuit-QED setups.
	Therefore, 
	the observation of the predicted novel mechanisms is within current 
	experimental 
	reach.
As an outlook, this study opens a new research direction 
in 
exploring exotic phenomena in nonlinear waveguide QED. 
In the future,  it is 
intriguing to
consider the coupling of emitter 
pairs to topological 
waveguides~\cite{Ling2024}, and investigate its nonlinear chiral quantum 
optics~\cite{Wang2024}.

 \section*{Acknowledgments}
 
 The quantum dynamical simulations are based 
 on open source code 
 QuTiP ~\cite{Johansson12qutip,Johansson13qutip}.  
 X.W.~is supported by 
 the National Natural Science
 Foundation of China (NSFC; No.~12174303), and the 
 Fundamental 
 Research Funds for the Central Universities (No. xzy012023053). T.L. acknowledges the 
 support from National Natural Science Foundation of China (Grant No.~12274142), the Fundamental Research Funds for the Central Universities (Grant No.~2023ZYGXZR020), Introduced Innovative Team Project of Guangdong Pearl River Talents Program (Grant
 No. 2021ZT09Z109), and the Startup Grant of South China University of Technology (Grant No.~20210012). A.M. was supported by the 
 Polish National Science Centre (NCN)
 under the Maestro Grant No. DEC-2019/34/A/ST2/00081. F.N. is supported in 
 part by:
 Nippon Telegraph and Telephone Corporation (NTT) Research,
 the Japan Science and Technology Agency (JST)
 [via the CREST Quantum Frontiers program Grant No. JPMJCR24I2,
 the Quantum Leap Flagship Program (Q-LEAP), and the Moonshot R\&D Grant 
 Number JPMJMS2061],
 and the Office of Naval Research (ONR) Global (via Grant No. 
 N62909-23-1-2074).

\appendix
\section{Properties of the band-edge modes $K= \pm \pi/2$} 
\label{A}

\begin{figure}[tbph]
	\centering \includegraphics[width=8.5cm]{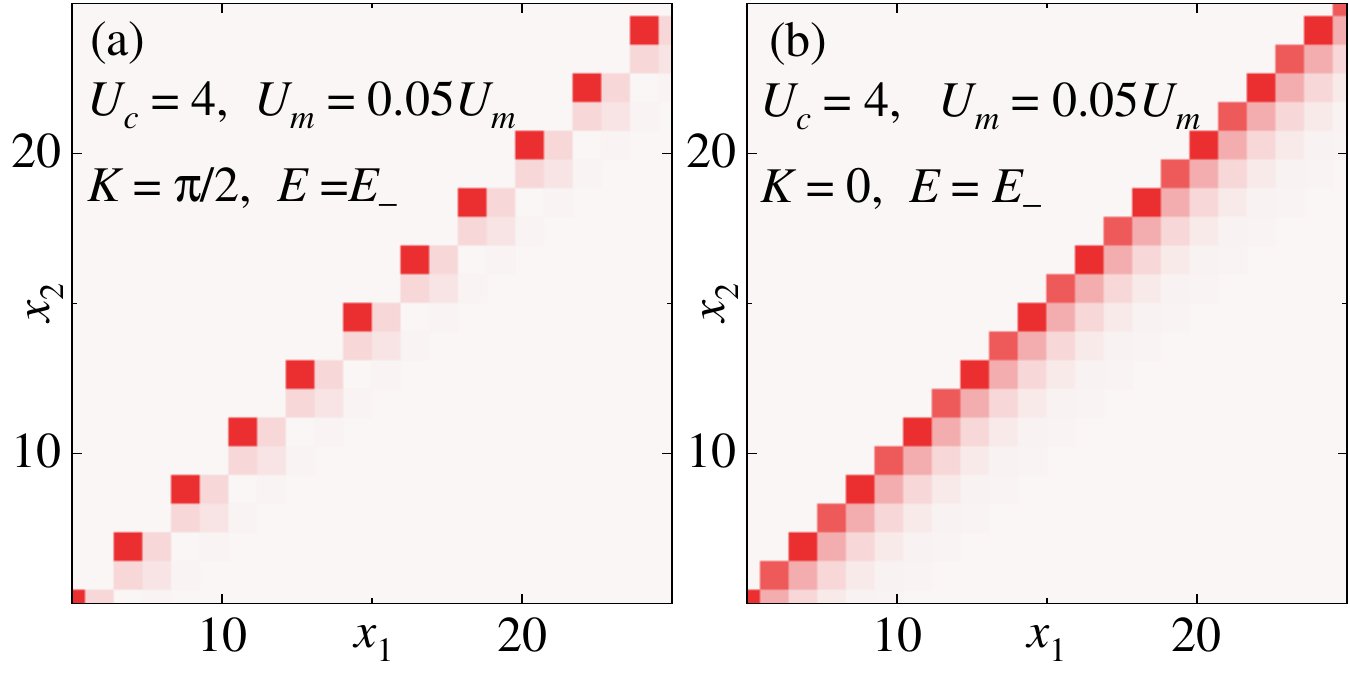}
	\caption{Wavefunctions $\Psi \left(x_c,r 
		\right)$ of the modes in the lower band at 
		(a) 
		$K=\pi/2$ and (b) $K=0$, 
		respectively. The waveguide parameters are the same 
		as those in 
		Fig.~2 in the main text.}
	\label{fig1s}
\end{figure}

In Fig.~2 of the main text, by considering a 
finite waveguide with periodic boundary conditions, we show the two-photon 
spectrum obtained 
by diagonalizing a nonlinear waveguide.  Notably, we observe 
the emergence of a band gap within the doublon spectrum, and its width 
increases proportionally to the modulation strength $U_m$. The situation 
is similar to a conventional photonic crystal waveguide, where the 
eigenwavefunction at the band edge of is localized on sites possessing a 
low (high) refractive index. The distribution properties 
of the field are led by the destructive interference resulting from 
multiple reflections.
Similarly, the behavior of doublons is also influenced by the periodic 
nonlinearity, and the characteristics of its wavefunction at the band 
edge resemble those of a single-photon crystal waveguide.

To substantiate the above discussions, we proceed to 
analyze the characteristics of the modes at $K_0=\pi/2$. Concerning  
the down/up energy levels, the eigenenergy is $$E_\mp=-\sqrt{\left( U_c\pm 
	U_m \right)
	^2+8J^2},$$ and the ratio in Eq.~(\ref{fk01})
is derived as
\begin{equation}
\frac{\psi _{K_0}^{(0)}\left( r=0 \right)}{\psi _{K_0}^{(1)}\left( r=0 
	\right)}=\frac{\frac{U_m}{U_c\pm U_m}}{1-\frac{U_c}{U_c \pm U_m}}=\pm 1.
\label{phi_K_0}
\end{equation}
Moreover, according to Eq.~(\ref{FKpi}), the following relation is valid
\begin{equation} 
f_{K}^{(0)}\left( E_\mp,r \right) =f_{K}^{(1)}\left( E_\mp,r \right). 
\end{equation}
For the down/up energy level at $K_0$,  
$f_{K}^{(0)}\left( 
E_\mp,r \right)$ derived as 
\begin{align}
&f_{K}^{(0)}\left( E_\mp,r \right)=\int{\frac{e^{iqr}}{E_\mp+4J\cos \left( 
		\frac{K}{2} \right) \cos 
		q}dq}\ \notag \\
	&\propto
\left( \frac{\sqrt{\left( U_c\pm U_m \right) ^2+8J^2}-\left( U_c\pm U_m 
	\right)}{2\sqrt{2}J} \right) 
^r \notag \\
&=\exp\left[ -\frac{r}{L_c^\mp(K_0)}\right],
\label{fkE0}
\end{align}
where 
\begin{align}
\frac{1}{L_c^{\mp}(K_0)}&=\ln \frac{2\sqrt{2}J}{\sqrt{\left( U_c\pm U_m \right) ^2+8J^2}-\left( U_c\pm U_m \right)} \notag 
\\
&=\,\,\ln \frac{2\sqrt{2}J}{-E_{\mp}-\sqrt{E_{\mp}^{2}-8J^2}} \notag 
\end{align}
is the decay length describing the joint probability of detecting two photons 
at the 
positions separated a distance $r$.

Equation~(\ref{fkE0}) implies that the doublon wavefunction decays as 
the separation distance between two photons increases. 
By substituting Eqs.~(\ref{phi_K_0})-(\ref{fkE0}) into Eq.~(\ref{phiKr}), one 
can find the wavefunctions at $K_0=\pm \pi/2$ for the lower/upper energy 
levels
\begin{align}
&\Psi_{K_0} \left(x_c,r \right) = \frac{1}{\sqrt{N }}e^{iK_0x_c}\left[ \psi 
_{K_0}^{0}\left( r \right) + e^{-i\pi x_c} \psi _{K_0}^{1}\left( r 
\right) \right]\ \notag \\
&= \frac{1}{\sqrt{N} \psi_0}e^{iK_0x_c}\left( 1\pm e^{-i\pi 
	x_c}
\right) \exp \left[-\frac{r}{L_c^\mp(K_0)}\right],
\end{align}
where $\psi_0$ is the normalized factor.

Now we summarize the characteristics of the doublon wavefunction around 
the band 
edge: First, due to the localized nature of the nonlinearity at each 
site, the maximum amplitude of $\Psi(E_{\mp},x_c,r)$ occurs at $r=0$ and 
rapidly diminishes as $r$ increases [along the diagonal in 
Fig.~\ref{fig1s}(b)].
Second, the wavefunction exhibits periodic localization at sites with 
lower (higher) nonlinearity owing to the destructive interference of the 
multiple reflections. For the modes which are significantly distant from the 
band gap, the interference effect is weak due to the frequency and 
wave-vector mismatch, and the wavefunction distributes on both even 
and 
odd sites [see Fig.~\ref{fig1s}(b)].

\section{Dynamics for emitter pairs inside the band gap}

\subsection{Equations of motion}
\label{B1}
In this section, we derive the motion equation.  
By substituting Eq.~(\ref{Hint_m},\ref{Tstates_m}) into Schrödinger Equation, we obtain the following coupled differential equations: 
\begin{widetext}
\begin{gather}
i\frac{dc_e\left( t \right)}{dt}=\frac{g}{\sqrt{N}}\sum_k{\left[ c_{1k}\left( t \right) 
	e^{ikn_2}+c_{2k}\left( t \right) e^{ikn_1} \right]}, \label{cet0}
\\
i\frac{dc_{1k}\left( t \right)}{dt}=\delta _kc_{1k}\left( t \right) 
+\frac{g}{\sqrt{N}}c_e\left( t \right) e^{-ikn_2}+g\sum_K{M\left( K,k,n_1 
	\right)}c_K\left( t 
\right), \label{cit1}
\\
i\frac{dc_{2k}\left( t \right)}{dt}=\delta _kc_{2k}\left( t \right) 
+\frac{g}{\sqrt{N}}c_e\left( t \right) e^{-ikn_1}+g\sum_K{M\left( K,k,n_2 
	\right)}c_K\left( t 
\right), \label{cit2}
\\
i\frac{dc_K\left( t \right)}{dt}=\Delta _Kc_K\left( t \right) +g\sum_k{M^*\left( K,k,n_1 
	\right)}c_{1k}\left( t \right) +g\sum_k{M^*\left( K,k,n_2 \right)}c_{2k}\left( t 
\right), 
\label{cKt0}
\\
\begin{align}
M\left( K,k,n \right) &=\langle k,n|\Psi _K\rangle 
=\langle 0|\frac{1}{\sqrt{N}}\sum_m{e^{-ikm}a_m}a_n\Psi _K\left( m',n' \right) a_{m'}^{\dagger}a_{n'}^{\dagger}|0\rangle 
\notag \\
&=\frac{\sqrt{2}}{N}\sum_m{e^{-ikm}e^{iK\left( n+m \right) /2}\,\,u_K\!\left( \frac{m+n}{2},n-m \right)},
\end{align}
\end{gather}
\end{widetext}
where $\delta_k=\omega_k-\omega_q$ with $\omega_k=-2J\cos k$ being the 
single-photon
spectrum, and 
$\Delta_K=E_K-2\omega_q$. 
In our discussion, $c_{ik}$ are the amplitudes of the single-photon 
intermediate 
states,
which are extremely small due to the large detuning relation $\delta_k\gg 
g$.  
Consequently, one can adiabatically eliminate $c_{ik}(t)$ by assuming its 
evolution to be time-independent. By setting $\dot{c}_{ik}\left( t 
\right)=0$, 
Eqs.~(\ref{cit1})-(\ref{cit2}) result in
\begin{equation}
c_{jk} ( t ) \! = \! -\frac{g}{\delta _k} \! \left[ \! \frac{e^{-ikn_j}}{\sqrt{N}}c_e\left( t \right) \!+\! \sum_K{M( K,k,n_j)}c_K( t ) \! \right]\!\!,
\label{cik_t}
\end{equation}
where $j=1,2$.
Altogether, by substituting Eq.~(\ref{cik_t}) into Eqs.~(\ref{cet0}) and (\ref{cKt0}), 
we end up with the coupled equations Eq.~(\ref{coupled_eq1}), in the main text.

In our work, we assume that the nonlinearity $U_c$ is large and the doublon 
spectum is 
well-separated from the scattering state.
When the frequency of the emitter pair resides within the doublon band 
gap, both emitters are unable to completely release their energy into the 
nonlinear waveguide via supercorrelated emission channels. The 
probability $|c_e(t)|^2$ undergoes a phenomenon known as \emph{fractional 
	decay}. In other words, the excitation partially dissipates into the 
waveguide while also remaining localized within the two emitters. 
\begin{figure}[tbph]
	\centering \includegraphics[width=8.5cm]{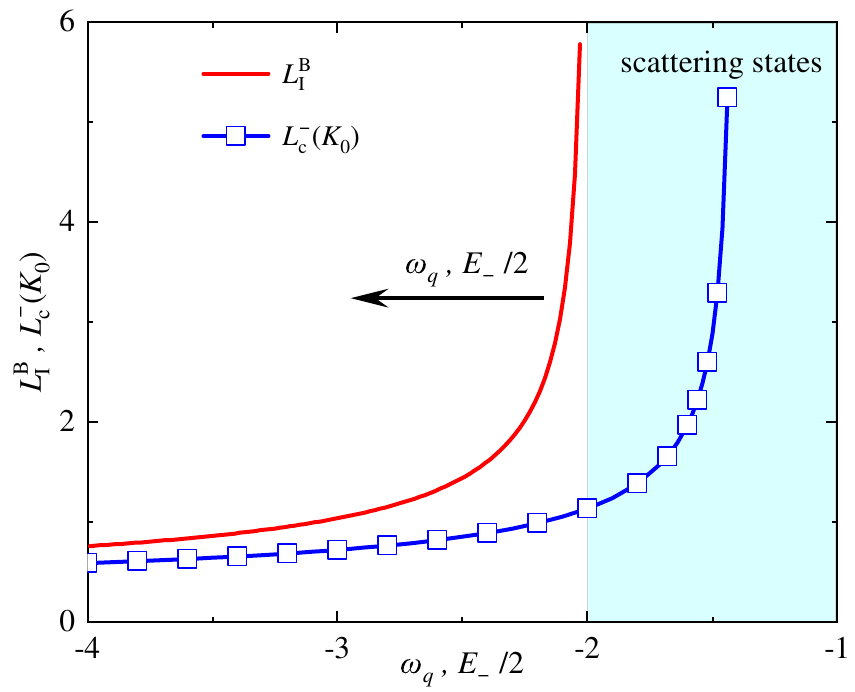}
	\caption{Decay length $L_{\mathrm{I}}^{\text{B}}$ of the SPBS (the 
		correlation length $L_c^{-}(K_0)$ for the DBS
		mode
		$K_0$), versus the emitter frequency $\omega_q$ (mode frequency 
		$E_-/2$). The 
		waveguide 
		parameters 
		are the 
		same as those in 
		Fig.~3 in the main text.}
	\label{fig2s}
\end{figure}

Moreover, compared with the single-photon case, the field distribution on the 
waveguide is 
much more complex. First, because the single-emitter frequency $\omega_q$ 
lies 
outside 
of the  
single-photon band structure, i.e, $\omega_q<-2J$, there are
single-photon bound 
states (SPBSs). Second, the coupled differential 
equations~(\ref{coupled_eq1}) 
and  
(\ref{coupled_eq2}) indicate that a doublon bound state (DBS) containing two 
strong-correlated photons also exists in this system. In the following, we  
derive the 
field distributions for these two kinds of bound states.

\subsection{Single-photon bound state}
\label{B2}
There are SPBSs
originating from individual emitters. 
Their existences can be explicitly confirmed by analyzing Eq.~(\ref{cik_t}), 
which approaches a steady state in the long-time limit, leading to the 
generation of SPBSs.
Considering the example of SPBS seeded by emitter 2, the 
steady-state population of the photonic component becomes 
\begin{align}
&\lim_{t\rightarrow \infty} c_{1k}\left( t \right) 
\notag \\
&=-\lim_{t\rightarrow \infty} \frac{g}{\delta _k}\left[ \frac{1}{\sqrt{N}}c_e\left( t \right) e^{-ikn_2}+\sum_K{M\left( K,k,n_1 \right)}c_K\left( t \right) \right] 
\notag \\
&\simeq -\lim_{t\rightarrow \infty} \frac{g}{\delta _k}\frac{1}{\sqrt{N}}e^{-ikn_2}c_e(t).
\end{align}
In our discussion, both 
the DBS and SPBS are only weakly excited. The excitations are mostly 
trapped 
inside 
the emitters, and we can approximate
$c_K(t)\ll c_e(t)\simeq 1$. Therefore, the SPBS wavefunction is 
\begin{align}
\psi_{2}(n)&=\frac{1}{2\pi}\int_{-\pi}^{\pi}{\frac{g}{w_e+2J\cos \left( k 
		\right)}  e^{ikn}e^{-ikn_2}dk}\ \notag \\
	&\simeq\ 
\frac{g}{2\pi}\int_{-\pi}^{\pi}{\frac{e^{ik\left( n-n_2 
			\right)}}{w_e+2J\cos 
		\left( k 
		\right)}dk}. \label{SPBS2}
\end{align}
The SPBS produced by the emitter 
1, i.e., 
$\psi_{1}(n)$, can also obtained by replacing $n_2\rightarrow n_1$ in  
Eq.~(\ref{SPBS2}).
The SPBS for the emitter $i$ can be written as
\begin{equation}
\psi_{i}(n)\simeq  \frac{gJ}{\sqrt{\left( w_e 
		\right) 
		^2-4J^2}}\exp\left(-\frac{|n-n_i|}{L_{\mathrm{I}}^{\text{B}}}\right),
\end{equation}
where the decay length $L_{\mathrm{I}}^{\text{B}}$ is 
\begin{equation}
L_{\mathrm{I}}^{\text{B}}=-\left(\ln \frac{-w_e-\sqrt{w_e 
		^2-4J^2}}{2J} \right)^{-1}.
\end{equation}

In Figure \ref{fig2s}, we plot $L_{\mathrm{I}}^{\text{B}}$ 
as a function of the single emitter frequency $\omega_q$, which shows 
that with increasing $\omega_q$, the decay length of the SPBS decreases 
rapidly. At $\omega_q=-2.5J$, the decay length is approximately 
$L_{\mathrm{I}}^{\text{B}}\simeq 1$.
Therefore, when $\delta_{\mathrm{I}}$ is large, the SPBS is strongly 
localized 
around the coupling point.

\subsection{Doublon bound state}
\label{B3}

We derive the DBS wavefunction. Given 
that two emitters are initially excited, we rewrite the coupled 
differential Eqs.~(\ref{coupled_eq1})-(\ref{coupled_eq2}) in the
Laplace space, i.e., 
\begin{gather}
s\tilde{c}_e\left( s \right) -1=i\frac{1}{\sqrt{N}}\sum_K{\mathcal{G} _K\left( n_1,n_2 \right) \tilde{c}_K\left( s \right)},
\\
\left( s+i\Delta _K \right) \tilde{c}_K\left( s \right) =i\frac{1}{\sqrt{N}}\mathcal{G} _{K}^{*}\left( n_1,n_2 \right) \tilde{c}_e\left( s \right) .
\end{gather}
The above equations give
\begin{gather}
\tilde{c}_e\left( s \right) =\frac{1}{s+\Sigma _e(s)}, \notag 
\\
\Sigma _e(s)=\frac{1}{N}\sum_K{|\mathcal{G} _K\left( n_1,n_2 \right) |^2\frac{1}{\left( s+i\Delta _K \right)}},
\end{gather}
where $\Sigma_e(s)$ can be interpreted as the self-energy. The real-time 
evolution is recovered as 
\begin{equation}
c_e\left( t \right) \! = \!\frac{1}{2\pi i}\underset{E\rightarrow \infty}{\lim}\int_{\epsilon 
	-iE}^{\epsilon +iE}{\frac{1}{s+\Sigma_e(s)} e^{st}ds}, \quad \epsilon>0.
\label{laplace_ce}
\end{equation}
Note that the emitter-pair frequency is quite close to the lower band edge of 
the 
doublon.
Therefore, only the modes around $K_0=\pi/2$ are excited with high 
probabilities.

\begin{figure}[tbph]
	\centering \includegraphics[width=8cm]{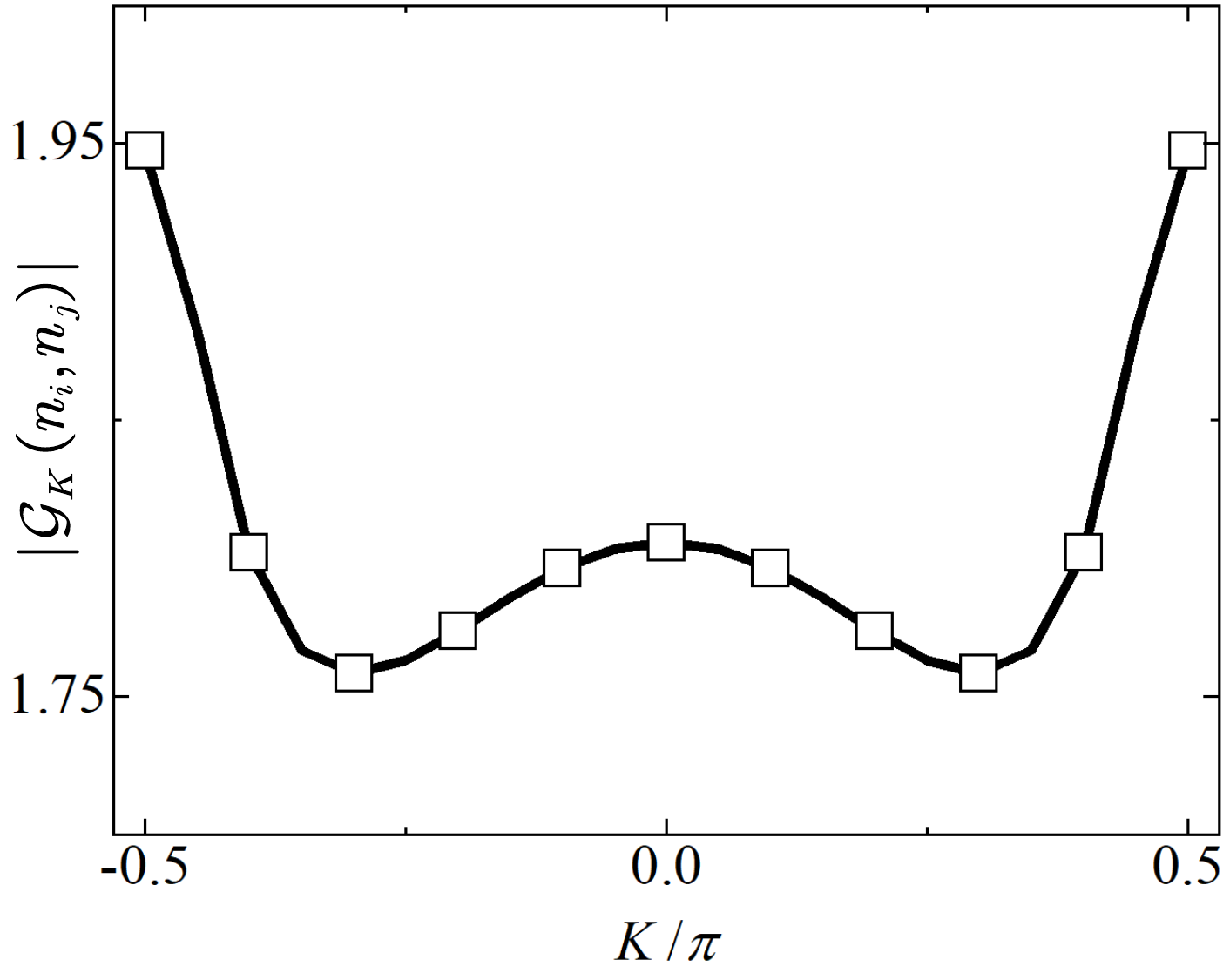}
	\caption{Amplitude of the coefficient $\mathcal{G}_K( n_i,n_j)$ versus 
	the 
	wave vector 
		$K$ with $g=1$. Here we set $n_i=n_j=0$. The 
		parameters 
		are the same as those in Fig.~3 used in the main text.}
	\label{fig3s}
\end{figure}

Before moving forward, we conduct a numerical analysis of the properties 
of the coefficient $\mathcal{G}_K\left( n_i,n_j \right)$. As depicted in 
Fig.~\ref{fig3s}, where $1.75<|\mathcal{G}_K\left( n_i, n_j \right)|<1.95$ 
($n_i=n_j=0$), we observe 
that $\mathcal{G}_K\left( n_i, n_j \right)$ varies within a narrow range 
versus $K$.
Therefore, we can approximate it independent of $K$ with
$\mathcal{G}_{K}\left( 
n_1,n_2 \right)\simeq \mathcal{G}_{K_0}\left( 
n_1,n_2 \right)$.
Moreover, we assume that
the dispersion relation around the edge of the lower band 
is described by a quadratic form, i.e., $\Delta _K\simeq  
\delta_{\mathrm{II}} 
+\alpha 
\left( 
K-\pi /2 \right) ^2 $ around $K=\pi/2$. By replacing the summation over 
$K$ with an integral, 
the self energy can be written as 
\begin{align}
\Sigma _e(s)&\simeq \frac{|\mathcal{G} _{K_0}\left( n_1,n_2 \right) |^2}{\pi}      \!\!    \int_0^{\pi} \!\!dK\! \frac{1}{s+i\left[ \delta _{\mathrm{II}}+\alpha \left( K-\pi /2 \right) ^2 \right]}
\notag \\
&\simeq
|\mathcal{G}_{K_0}\left( n_1,n_2 \right) |^2\frac{1}{\sqrt{\left( is-\delta_{\mathrm{II}} 
		\right) 
		\alpha}}, \label{sigeself}
\end{align}
where we have extended the integral bound to infinity. 
By substituting Eq.~(\ref{sigeself}) into Eq.~(\ref{laplace_ce}), we derive 
the steady-state population for $c_e(t)$ via the residue theorem, i.e.,
\begin{gather}
\lim_{t\rightarrow \infty} |c_e\left( t \right) |^2 =|\mathrm{Res}\!\left( s_0 \right) |^2,
\end{gather}
where $s_0$ is the unique pure imaginary pole for the denominator of 
$\tilde{c}_e\left( s \right)$, 
i.e., $s_0+\Sigma_e(s_0)=0$, and
\begin{align}
\mathrm{Res}\left( s_0 \right) \! &=\! \left. \frac{1}{1+\partial _s\Sigma _e\left( s \right)} \right|_{s=s_0}
\notag \\
&= \! \frac{1}{1-\frac{\alpha}{2}|\mathcal{G} _{K_0}\left( n_1,n_2 \right) |^2 \left[ \left( s_0+i\delta _{\mathrm{II}} \right) \alpha \right] ^{-\frac{3}{2}}}.\!\!\!\!\!
\end{align}
By following the process of evaluating non-Markovian 
dynamics in Sec.~II, one can also analysis non-Markovian to Markovian 
transition for this supercorrelated decay process.
Now we derive the steady field distribution in the waveguide, i.e., 
the DBS.
In the long-time limit $t\rightarrow \infty$, we set $\dot{c}_K\left( t 
\right)=0$ 
in Eq.~(\ref{coupled_eq2}). 
The steady amplitude for 
mode $K$ is 
\begin{equation}
\lim_{t\rightarrow \infty} c_K\left( t \right) =\frac{1}{\sqrt{N}\Delta _K}\mathcal{G} _{K}^{*}\left( n_1,n_2 \right) \mathrm{Res}\left( s_0 \right) .
\end{equation}

Be denoting $x_m=\left( n_1+n_2 \right) /2$ as the center position of the 
two emitters, the wavefunction of the DBS can be expressed as follows:
\begin{align}
&\Psi _d(x_c,r)=\sum_K{c_K}\left( t_{\infty} \right) \frac{1}{\sqrt{N}\psi _0}e^{iKx_c}u_K\left( x_c,r \right) 
\notag \\
&=\frac{\mathrm{Res}\left( s_0 \right)}{\pi \psi _0}\int_0^{\pi} \mathcal{G} _{K}^{*}\left( n_1,n_2 \right) u_K\left( x_c,r \right) \frac{e^{iK\left( x_c-x_m \right)}}{\Delta _K} {dK}
\notag \\
&\simeq \! \! \frac{\mathrm{Res}\left( s_0 \right)}{\pi \psi _0}\mathcal{G} _{K_0}^{*} ( n_1,n_2 ) u_{K_0} ( x_c,r) \! \! \int_{-\infty}^{\infty} \! \frac{e^{i\delta K(\! x_c-x_m \!)}}{\delta _{\mathrm{II}} \! + \! \alpha \delta K^2} {d\delta K},
 \label{psidoublon}
\end{align}
where we assume that 
$u_K\left( x_c,r \right)$ 
is also independent of $K$ since only the modes around $K_0=\pi/2$ are 
excited with high 
probabilities. Moreover, we approximate $\mathcal{G}_K\left( n_1,n_2 \right)\simeq 
\mathcal{G}_{K_0}\left( 
n_1,n_2 \right)$ as a constant in Eq.~(\ref{psidoublon}).
Finally, the wavefunction for the DBS is derived as
\begin{gather} \label{DBS_state_Res}
\Psi _d(x_c,r)=A_d\mathrm{Res}\left( s_0 \right) u_{K_0}\left( x_c,r \right) e^{-\frac{|x_c-x_m|}{L_{\mathrm{II}}^{\mathrm{B}}}}
, \\
A_d=\frac{\mathcal{G}_{K_0}^{*}\left( n_1,n_2 \right)}{\psi _0\sqrt{\delta _{\mathrm{II}}\alpha}},\qquad L_{\mathrm{II}}^{\mathrm{B}}=\sqrt{\frac{\alpha}{\delta _{\mathrm{II}}}}. \notag
\end{gather}
Given that the largest fraction of the energy is still trapped inside the 
emitters, we approximate 
$|\mathrm{Res}\left( s_0 \right)|\simeq 1$ and Eq.~(\ref{DBS_state_Res}) is simplified as Eq.~(\ref{DBS_state_m}), in the main text.

\section{Four-body interactions by exchanging doublons}
Let us now consider a scenario involving four emitters, forming two pairs 
that couple to the same waveguide (refer to Fig.~1 in the main text). We 
assume that the initial two excitations are localized within emitter 
1 and 2, i.e., 
$c_{12}^e(t_0)=1$, and the corresponding state is denoted as 
$|eegg\rangle$. In principle, the populations $c_{13}^e(t)$, 
$c_{23}^e(t)$, $c_{24}^e(t)$, and $c_{14}^e(t)$ are nonzero
due to exchanging a single photon when the SPBSs in different pairs overlap. 
For instance, 
through the exchange of one photon between the emitters 1 and 3, the 
single-photon transition $|eegg\rangle\rightarrow |egeg\rangle$ occurs. 
To observe high-fidelity four-body Rabi oscillations between $c_{12}^e(t)$ 
and $c_{34}^e(t)$, it is crucial to suppress undesired single-photon 
processes. This can be achieved by positioning the emitter pairs at a 
distance significantly greater than the decay length of SPBS, i.e., 
$$D_q=(n_3+n_4-n_1-n_2)/2\gg L_{\mathrm{I}}^{\text{B}}.$$
Under these conditions, the 
populations 
$$c_{13}^e(t),\ c_{23}^e(t),\ c_{24}^e(t),\ c_{14}^e(t)\simeq 0,$$
and the system evolution is reduced to
\begin{align}
i\dot{c}_{12}^{e}\! \left( t \right) 
\!&=\!-\frac{1}{\sqrt{N}}\sum_K{\mathcal{G}_{1K}c_K\left( t \right)},
\\
i\dot{c}_{34}^{e} \! \left( t \right) 
\!&=\!-\frac{1}{\sqrt{N}}\sum_K{\mathcal{G}_{2K}c_K\left( t \right)},
\\
i\dot{c}_K \! \left( t \right) \!& = \! \Delta _{K_d}c_K\left( t \right) 
\!-\!\! \frac{1}{\sqrt{N}} \! \left[ \mathcal{G}_{1K}^{*}c_{12}^{e} \!\left( 
t \right) \!+\!\mathcal{G}_{2K}^{*}c_{34}^{e}\!\left( t \right) 
\right],\!\!\!\!\!
\end{align}
where $$\mathcal{G}_{1K}=\mathcal{G}_K(n_1,n_2),\quad 
\mathcal{G}_{2K}=\mathcal{G}_K(n_3,n_4).$$
The doublon mode $K$ can mediate the coherent exchange of excitations between 
the 
two emitters. In our analysis, the doublon is virtually excited, allowing us 
to 
adiabatically eliminate its degree of freedom by assuming $\dot{c}_K\left( t 
\right) =0$.  
Finally, the coupled 
differential equations are reduced to
\begin{gather}
i\dot{c}_{12}^{e}\left( t \right)\! =\! -\frac{1}{N}\! \sum_K\! {\frac{\left[ 
|\mathcal{G}_{1K}|^2c_{12}^{e}( t ) 
+\mathcal{G}_{1K}\mathcal{G}_{2K}^{*}c_{34}^{e} ( t ) \right]}{\Delta _K}}, 
\label{couple_ee1}
\\ 
i\dot{c}_{34}^{e}\left( t \right)\! =\! -\frac{1}{N}\! \sum_K\! {\frac{\left[ 
|\mathcal{G}_{2K}|^2c_{34}^{e}( t ) 
+\mathcal{G}_{2K}\mathcal{G}_{1K}^{*}c_{12}^{e}( t ) \right]}{\Delta _K}}.
\label{couple_ee2}
\end{gather}
One can find that Eqs.~(\ref{couple_ee1}) and (\ref{couple_ee2}) correspond to an effective Hamiltonian Eq.~(\ref{Heff}), in the main text.

%

\end{document}